\documentclass[aps,prl,twocolumn,groupedaddress,showpacs]{revtex4}

\usepackage{graphics}
\begin{document}


% %%%%%%%%%%%%%%%%%%%%%%%%%%%%%%%%%%%
%\preprint{}

% %%%%%%%%%%%%%%%%%%%%%%%%%%%%%%%%%%%

\title{Order parameter node removal in the \textit{d-wave} superconductor $YBa_{2}Cu_{3}O_{7-x}$ ~{under magnetic field}}



\author{R. Beck}
\email[]{Roy@tau.ac.il}
\homepage[]{www.tau.ac.il\~supercon}
\author{Y. Dagan}
\thanks{Present address: Center for Superconductivity Research,
Department of Physics, University of Maryland at College Park, College Park,
Maryland 20742}

\author{A. Milner}

\author{A. Gerber}

\author{G. Deutscher}

\affiliation{School of physics and Astronomy, Raymond and Beverly Sackler faculty of Exact Science, Tel-Aviv University, 69978 Tel-Aviv, Israel}

\date{\today}

\begin{abstract}
Whether the node in the order parameter characteristic of a $d-wave$
superconductor can or cannot be removed by an applied magnetic field has
been a subject of debate in recent years. Thermal conductivity results on
the high Tc superconductor $Bi_{2}Sr_{2}CaCu_{2}O_{8}$ originally
explained by Laughlin in terms of such a node removal were complicated by
hysteresis effects, and judged inconclusive. We present new tunneling data
on $YBa_{2}Cu_{3}O_{7-x}$ that support the existence of the node
removal effect, under specific orientations of the sample's surfaces and
magnetic field. We also explain the hysteretic behavior and other previous
tunneling results so far not understood satisfactorily, attributing them to
a combination of node removal and Doppler shift of low energy surface bound
states.
\end{abstract}

 \maketitle

The excitation spectrum of a conventional superconductor (Low Tc)
is characterized by an almost independent momentum ($s-wave$)
energy gap, $\Delta$. This is not the case in the high Tc cuprate
superconductors; there it is broadly agreed that the ground state
superconducting order parameter (OP) is strongly momentum
dependent, it is maximal in the direction of the crystallographic
axes a and b; in most cases it appears to have the pure
$d_{x^{2}-y^{2}}$ symmetry, being zero at 45$^{o}$ between these
axes (the node directions), where it changes sign. In contrast to
the finite energy $\Delta$ required to excite a low energy
quasi-particle in a Low Tc superconductor, such a quasi-particle
can be excited with an infinitely small energy in a $d-wave$
superconductor along the nodes. This is no longer the case if an
additional imaginary component is present in the OP, on that case
the energy spectrum of the superconductor is fully gapped.

Theoretically, it has been suggested that such an imaginary component can
result from instability of the $d-wave$ OP under perturbations such as surface
pair breaking \cite{b1}, impurities \cite{b2}, proximity effect \cite{b3, b4} and magnetic field
\cite{b5}. Another view is that a phase transition occurs at a certain doping
level \cite{b6,b7}, or magnetic field \cite{b8} from pure $d-wave$ to a nodeless OP having
the $d_{x^{2}-y^{2}}+id_{xy}$ or $d_{x^{2}-y^{2}}+is$ symmetry. The $id_{xy}$ component breaks
both time and parity symmetries, hence, as pointed out by Laughlin \cite{b8}, it
involves boundary currents that flow in opposite directions on opposite
faces of the sample. They produce a magnetic moment which, through
interaction with the magnetic field, lowers the free energy by a term
proportional to $B\cdot d_{xy}$, where B is the magnetic field. On the other hand, in
the zero temperature limit, node removal costs an energy proportional to
$\vert id_{xy}\vert^{3}$. Minimization of the sum of the two contributions
leads to an amplitude $\vert d_{xy}\vert = A\cdot B^{1/2}$, where A is a coefficient.

Experimentally, two sets of experiments have been interpreted as indicating
that a magnetic field, applied perpendicular to the $CuO_{2}$ planes, can
indeed induce a node less OP. Measurements of the thermal conductivity
\textit{$\kappa (H)$ }on $Bi_{2}Sr_{2}CaCu_{2}O_{8}$ (Bi2212) single crystals have shown a
decrease followed by a plateau at a certain field \cite{b9}. This field was
interpreted as that beyond which a finite \textit{id}$_{xy}$ component appears at the
finite temperature where the experiment is performed \cite{b8}. The finite gap in
the plateau region prevents the excitation of additional quasi particles.
Field hysteresis of the thermal  conductivity has however led to a
controversy as to the actual origin of the plateau, and this issue has
remained unresolved \cite{b10}.

A second set of experiments possibly indicating the occurrence of a nodeless OP is the field evolution of the conductance of in-plane tunnel
junctions formed at the surface of $YBa_{2}Cu_{3}O_{7-x}$ (YBCO)
films oriented perpendicular to the $CuO_{2}$ planes \cite{b23}. This conductance
presents a peak at zero bias (known as Zero Bias Conductance Peak, ZBCP),
which splits when the field is applied parallel to the film's surface, and
perpendicular to the $CuO_{2}$ planes. The ZBCP is one of the clearest
manifestations of an OP having nodes, because it comes about due to a change
of phase by $\pi $ upon reflection at a (110) surface, which generates low
energy surface states, or Andreev bound states \cite{b13} (the fact that ZBCPs are
also observed for other in-plane orientations is generally interpreted as
due to surface roughness \cite{b11}). Peak splitting may indicate node removal,
the new peak position, \textit{$\delta (H)$}, giving the amplitude of the $d_{xy}$ component \cite{b6}.
However, a different explanation of the peak splitting was also offered, in
terms of a Doppler shift of the Andreev bound states \cite{b11}. This Doppler
shift is equal to $\textbf{v}_{s}\cdot \textbf{p}_{F }$, where
$\textbf{v}_{s}$ is the superfluid velocity associated with the
Meissner currents, and $\textbf{p}_{F}$ the Fermi momentum of the
tunneling quasi-particles. There are difficulties with both interpretations.
The amplitude of the $\textit{id}_{xy}$ component should be essentially reversible at
the fields of interest (in the Tesla range), where the thin film samples are
well into the Bean complete penetration limit, while in fact a strong field
hysteresis is observed \cite{b18, b25, b12}. As for the Doppler shift, it should
vanish at film thickness smaller than the London penetration depth where the
Meissner superfluid velocity is much reduced, varying as $Vs=e\lambda H tanh(d/2\lambda )$, where $d$ is the
film thickness,$\lambda$ the penetration depth and $H$ is the magnetic field at the
sample surface; while in fact it is slightly changes at $d = (\lambda /2)$ \cite{b26}. A further
difficulty is that, for reasons that have remained unclear until now, field
splitting of the ZBCP is not always observed. For instance, it was not seen
in $La_{1.85}Sr_{0.15}CuO_{4}$ and YBCO grain boundary junctions \cite{b15}, nor in junctions prepared on
Bi2212 single crystals \cite{b16}. Thus, a consensus has not yet been reached as
to whether a field can induce a finite sub-gap in a \textit{d-wave} superconductor.

We present in this Letter a new series of tunneling experiments that clarify
the origins of the field splitting phenomenon, and establish the conditions
under which node removal occurs. Our central result is that node removal can
be observed in (110) oriented samples only. It is most clearly seen in
decreasing fields, for which the ZBCP splitting is not affected by Doppler
shift effects, and is thickness independent. Doppler shift does affect data
taken in increasing fields, which are thickness dependent. Hysteresis is due
to the properties of the Bean Livingston barrier, which is effective against
flux penetration in increasing fields, but not against flux exit in
decreasing fields. Data taken on (110) oriented samples in decreasing
fields, starting from a field of up to 16T, are in quantitative agreement
with Laughlin theory.

YBCO films nearly optimally doped, having thickness ranging from 600\AA up to
3200\AA, were prepared in the (110) and (100) orientations by the template
method, using $SrTiO_{3}$ and $LaSrGaO_{4}$ substrates of the appropriate
orientation \cite{b17,b18}. Critical temperatures of all films were in the range of
88K to 90K. Junctions were prepared by pressing In (Indium) pads on the
films fresh surface \cite{b17,b18}. All junctions were measured at 4.2K, and some
also at 1.6K. All junctions displayed an unsplit ZBCP in zero magnetic
field, irrespective of the film orientation. Junctions characterisitics were
measured as a function of field, applied parallel to the surface, either
parallel or perpendicular to the $CuO_{2}$ planes. A typical data set is shown
Fig.1 for a 600\AA , (110) oriented film. Measurements were systematically
taken in increasing and decreasing fields. Field splitting was only observed
when the field was applied perpendicular to the $CuO_{2}$ planes, in agreement
with previous results, confirming the uniaxial in-plane orientation of the
c-axis \cite{b17}. A total of 20 junctions were measured.

The peak position $\delta(H)$ measured on the 600$\AA$ thick, (110) oriented film whose
conductance characteristics are shown Fig.1, is plotted Fig.2a in increasing
and decreasing fields. In increasing fields, the peak position reaches about
4 meV at 5T. At higher fields, it cannot be determined because the peak
becomes too smeared, possibly because it merges with the main gap structure.
For thicker films, the rise in the peak position is faster. For a 3000$\AA$ thick film, it reaches 4 meV already at 3T. Data taken in decreasing fields
for the 600$\AA$ film starts at about 4 meV at 15T. By contrast with the
behavior in increasing fields, that in decreasing ones is thickness
independent.

The hysteresis amplitude is shown Fig.2b up to 5T, the highest field at
which it could be determined. Hysteresis saturates above 2T, a field of the
order of Hc. For thicker films, the hysteresis amplitude reaches larger
values, but the field dependence is similar. This hysteretic behavior can be
understood within the Doppler shift model of the ZBCP splitting \cite{b11}, if we
take into account the properties of the Bean Livingston barrier. This
barrier prevents flux entry up to fields of the order of Hc. Up to that
field, the superfluid velocity of the Meissner currents increases almost
linearly; beyond that field, it saturates. The Doppler shift, proportional
to the superfluid velocity, follows the same behavior. By contrast, in
decreasing fields, there is no barrier that prevents flux exit. This has
been shown theoretically by Clem \cite{b19}, verified experimentally by Bussiere
\cite{b20}, and shown also to apply in tunneling experiments by Moore and Beasley
\cite{b21}. Hence, when the field is decreased, the surface superfluid velocity
quickly reduces to zero, and so does the Doppler shift. This is the origin
of the hysteresis. The fact that the hysteresis amplitude increases with
film thickness is in line with this interpretation.

Our central result is shown Fig.3, which presents the ZBCP splitting
measured in decreasing fields for (110) oriented films having thickness
ranging from 3200$\AA$ down to 600$\AA$, plotted as a function of the square root of
the applied field. Data for all samples follow the law $\delta_\downarrow = A\cdot B^{1/2}$, with $A=1.1 meV/T^{1/2}$. Laughlin calculated for Bi 2212 a coefficient
$A=1.6 meV/T^{1/2}$. This coefficient is proportional to the square root
of the gap and that of the Fermi velocity. Taking into account that gap
values measured on Bi2212 (25 to 30 meV) are larger than those measured on
YBCO (17 to 20 meV), and assuming that values of the Fermi velocity are
similar for both compounds near optimum doping, there is a good quantitative
agreement between theory and experiment.

These results are quite different from those previously reported for (100)
\cite{b18} and (103) \cite{b27, b12, b25} oriented films. Generally speaking, field
splitting values are smaller for these orientations. For 3000$\AA$ (100)
oriented films, the splitting measured in increasing fields reaches only 1.5
meV at 5T \cite{b18}, as compared to 4 meV at 3T for a similar thickness in the
(110) orientation, as mentioned above. Differences are even larger in
decreasing fields, splitting values falling always well below the line shown
in Fig.3. For the 3000$\AA$ thick sample mentioned above, it is only of 0.5 meV
at 5T, as compared to 2.4 meV in the (110) orientation. In addition,
splitting values in the (100) orientation are strongly thickness dependent.
For samples thinner than 1600$\AA$, we find that they are too small to be
determined experimentally at 4.2K up to 6T. These results are consistent
with a splitting dominated in the (100) orientation by the Doppler shift effect, with the Laughlin
mechanism playing apparently no role.

Surface faceting is thought to be the reason for the ZBCP commonly
observed in (100) oriented films \cite{b11}. Recently, some direct
evidence has indeed been provided by STM measurements, suggesting
that (110) faces are present in films of that orientation
\cite{b17}. While the zero field ZBCP is similar for both
macroscopic orientations, their field splitting is entirely
different, as reported here. Our results demonstrate that in order
to observe a substantial splitting \textit{in decreasing fields
and in films smaller than the London penetration depth }-- namely,
under conditions where the Doppler shift effect is very weak --
one must use samples having the (110) orientation. Then, and only
then, the experimental data are in agreement with the law of
Laughlin, strongly suggesting that node removal does occur. Even
though (100) oriented films do have (110) facets at their outer
surface, the interface with the (100) LaSrGaO substrate and
PrBaCuO intermediate layer is presumably quite flat. We conjecture
that the absence of the second (110) surface prevents the flow of
Laughlin's currents on that face, making the establishment of the
\textit{id}$_{xy}$ component energetically unfavorable.

In conclusion, the presented tunneling data is consistent with
node removal in the $d-wave$ superconductor YBCO under magnetic
field if, and only if, the sample's boundaries have the (110)
orientation. As far as we know, this has not been predicted
theoretically. Previously not well understood experimental
results, such as the difference in field splitting between (110)
and (100) oriented films and its hysteretic behavior, can now be
explained. The absence of the ZBCP field splitting in grain
boundary junctions \cite{b15} can be understood as a combination
of two factors: first, the field being applied perpendicular to
the surface, vortices penetrate at low fields, and there cannot be
any substantial Doppler shift. Second, the geometry of the
boundaries is unfavorable for the flow of Laughlin's currents. The
same applies to junctions produced at the edge of Bi2212 crystals
\cite{b16} with the field being applied perpendicular to the
surface. It could be that the contradictory results reported in
thermal conductivity experiments, concerning the existence of a
field induced gap, also stem from the ability or inability of the
samples to carry boundary currents under the specific experimental
conditions.

\section{}
\begin{acknowledgments}
We are indebted to Malcom Beasley for pointing out to one of us
(G.D.) reference \cite{b21} on the effects of Bean Livingston
barrier as seen in a tunneling experiment, to Dr. H. Castro and A.
Kohen for helpful conversations. This work was supported by the
Heinrich Herz-Minerva Center for High Temperature
Superconductivity, by the Israel Science Foundation and by the
Oren Family Chair of Experimental Solid State Physics.
\end{acknowledgments}

%\bibliography{noderemoval}

\begin{references}
\bibitem{b1} Y. Tanuma and Y. Tanaka Y and S. Kashiwaya,Phys. Rev. B {\bf 64}, 214519 (2001).
\bibitem{b2} A.V. Balatsky and M.I. Salkola and A.Rosengren,Phys. Rev. B {\bf 51}, 15547 (1995).
\bibitem{b3} Y. Ohashi,Phys. Soc. Jpn. {\bf 65}, 823 (1996).
\bibitem{b4} A. Kohen and G. Deutscher,J. Low Temp. Phys (to be published).
\bibitem{b5} A.V. Balatsky,Phys. Rev. B {\bf 61}, 6940 (2000).
\bibitem{b6} Y. Dagan and G. Deutscher,Phys. Rev. Lett. {\bf 87}, 177004 (2001).
\bibitem{b7} M. Vojta and Y. Zhang and S. Sachdev,Phys. Rev. Lett. {\bf 85}, 4940 (2000).
\bibitem{b8} R.B. Laughlin,Phys. Rev. Lett. {\bf 80}, 5188 (1998).
\bibitem{b9} K. Krishana and N.P. Ong and Q. Li and G.D. Gu and N.Koshizuka,Science {\bf 277}, 83 (1997).
\bibitem{b10} A. Aubin, K. Behnia, Science {\bf 280}, 9a (1998).
\bibitem{b11} M. Fogelstr\"{o}m and D. Rainer and J. A. Sauls, Phys. Rev. Lett. {\bf 79}, 281 (1997).
\bibitem{b12} M. Covington \textit{et al.}, Phys. Rev. Lett. {\bf 79}, 277 (1997).
\bibitem{b13} C. R. Hu, Phys. Rev. Lett. {\bf 72}, 1526 (1994).
\bibitem{b14} S. Kashiwaya and Y. Tanaka and M. Koyanagi and H. Takashima and K. Kajimura, Phys. Rev. B {\bf 51}, 1350 (1995).
\bibitem{b15} L. Alff, Eur. Phys. J. B. {\bf 5}, 423 (1998).
\bibitem{b16} H. Aubin and L. H. Greene, Phys. Rev. Lett. {\bf 89}, 177001 (2002).
\bibitem{b17} Y. Dagan and R. Krupke and G. Deutscher, Phys. Rev. B {\bf 62}, 146 (2000).
\bibitem{b18} R. Krupke and G. Deutscher, Phys. Rev. Lett. {\bf 83}, 4634 (1999).
\bibitem{b19} J.R. Clem, n Proceeding of Thirteenth International Conference on Low Tempreture Physics Boulder, Colo., 1972m edited by K. D. Timmerhaus, W. J. O'Sullivan and E.F. Hammel, Vol. 3 (Plenum press, New York, 1974), p. 102.
\bibitem{b20} J.F. Bussiere, Phys. Lett. {\bf 58A}, 343 (1976).
\bibitem{b21} D.F. Moore and M.R. Beasley, Appl. Phys. Lett. {\bf 30}, 494 (1977).
\bibitem{b22} A. Sharoni \textit{et al.}, Phys. Rev. B {\bf 65}, 134526 (2002).
\bibitem{b23} G. Deutscher and Y. Dagan and A. Kohen and R. Krupke, Phyisca C {\bf 341-348}, 1629 (2000).
\bibitem{b25} M. Aprili and E. Badica and L. H. Greene, Phys. Rev. Lett. {\bf 83}, 4630 (1999).
\bibitem{b26} Y. Dagan and G. Deutscher, Phys. Rev. B {\bf 64}, 92509 (2001).
\bibitem{b27} J. Lesueur and L. H. Greene and W.L. Feldmann and A. Inam, Phyisca C {\bf 191}, 325 (1992).



\end{references}

\begin{figure}
\resizebox{!}{5cm}{\includegraphics{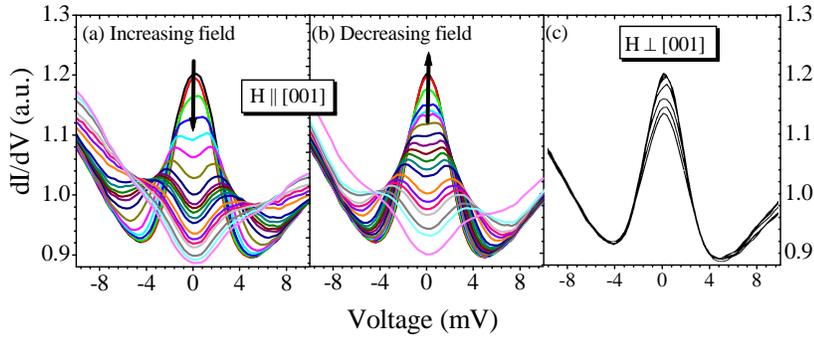}}
\caption{\label{FIG. 1} Normalized dynamical conductance G=dI/dV
Vs bias V for increasing (a) and decreasing (b) applied magnetic
fields for YBCO (110) oriented film at 4.2K. Film characteristics:
Tc= 88K, film thickness d=600\AA. The splitting ($\delta$) is
defined as half of the distance between the positions of the
conductance maxima. In increasing field it can be determined
clearly from field of about 0.1T up to 5T, and in decreasing
fields from 13T down to 0.9T. Applied fields (in Tesla) : 0, 0.1,
0.3, 0.5, 0.7, 0.9, 1.2, 1.5, 1.8, 2.1, 2.5, 3.0, 3.5, 5, 6, 7,
11, 13, 15. (c) behavior of the same junction for magnetic field
applied parallel to the $CuO$ planes at fields (in Tesla): 0, 0.5,
1, 2, 4, 8, 12 ,15.5. The strong anisotropy of the field effect
confirms the good in-plane orientation of the c-axis.}
\end{figure}

\begin{figure}
\includegraphics{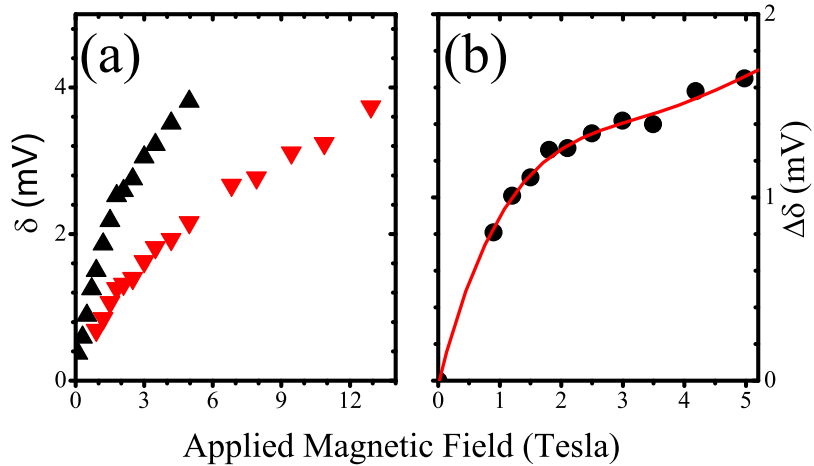}
\caption{\label{FIG. 2} Comparison between the field splitting
hysteresis curves for (110) and (100) in the presence of
increasing ($\bigtriangleup$) and decreasing ($\bigtriangledown$)
external magnetic fields. (a) ZBCP splitting ($\delta$) and (b)
ZBCP splitting difference $(\Delta \delta =\delta _{ \uparrow }-
\delta _{ \downarrow })$ for (110) 600$\AA$ thickness films at
4.2K. The line in (b) is a guide to the eye.}
\end{figure}

\begin{figure}
\includegraphics{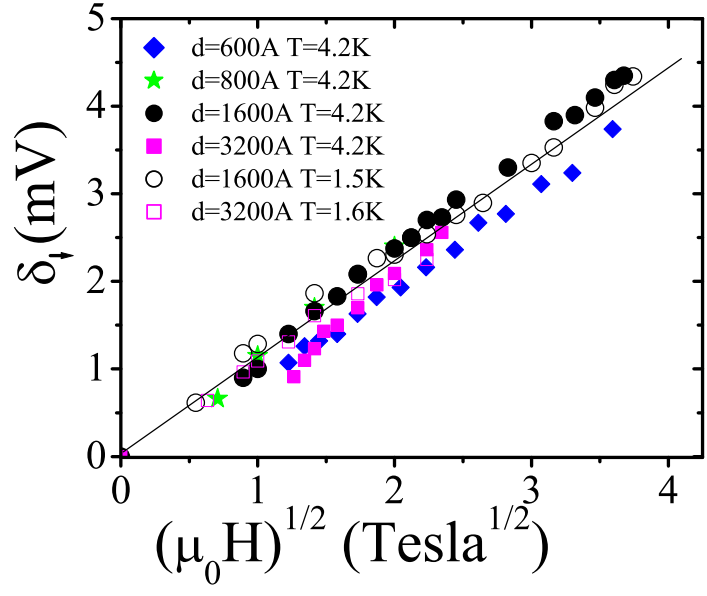}
\caption{\label{FIG. 3} ZBCP Field splitting measured in
decreasing fields for film thickness ranging from 3,200{$\AA$} to
600{$\AA$} as a function faceting square root of applied magnetic
field (H). The line is a linear fit to all points with
$1.1mV/T^{1/2}$ slope.}
\end{figure}

\end{document}